\documentclass[letterpaper, 10 pt, conference]{ieeeconf}  

\IEEEoverridecommandlockouts                              

\usepackage{graphicx}
\graphicspath{{figures/}}
\usepackage{algorithmic}
\usepackage{algorithm}
\usepackage{booktabs} 
\usepackage{subcaption}
\usepackage{amsmath} 
\usepackage{amssymb}  
\usepackage{bbm}
\usepackage[T1]{fontenc}
\usepackage{amsmath}
\usepackage{multirow}
 
\usepackage{url}
\usepackage{color}
\newcommand{\pa}{\textcolor{black}}
\newcommand{\prv}{\textcolor{black}}
\usepackage{todonotes}

\title{\LARGE \bf
 Deep Reinforcement Learning-based Intelligent Traffic Signal Controls with Optimized CO2 emissions  
}

\author{Pedram Agand$^{1}$, Alexey Iskrov$^{2}$, and Mo Chen$^{1}$
\thanks{This work received support from Breeze Traffic Inc. and the Mitacs Accelerate Program. We thank Alexander Kurtynin at Breeze Traffic Inc. for insightful discussions.}
\thanks{Disclaimer: This work has been accepted for publication in the International Conference on Intelligent Robots and Systems (IROS). © 2023 IEEE.  Personal use of this material is permitted.  Permission from IEEE must be obtained for all other uses, in any current or future media, including reprinting/republishing this material for advertising or promotional purposes, creating new collective works, for resale or redistribution to servers or lists, or reuse of any copyrighted component of this work in other works.}
\thanks{$^{1}$P. Agand and M. Chen are with Simon Fraser University, Burnaby, Canada
        {\tt\small \{pagand, mochen\}@sfu.ca}}%
\thanks{$^{2}$A. Iskrov is with Breeze Traffic Inc., Vancouver, BC, Canada
        {\tt\small  alexey@breezetraffic.com}}%
}
\begin{document}

\maketitle
\thispagestyle{empty}
\pagestyle{empty}

\begin{abstract}
Nowadays, transportation networks face the challenge of sub-optimal control policies that can have adverse effects on human health, the environment, and contribute to traffic congestion. Increased levels of air pollution and extended commute times caused by traffic bottlenecks make intersection traffic signal controllers a crucial component of modern transportation infrastructure. Despite several adaptive traffic signal controllers in literature, limited research has been conducted on their comparative performance. Furthermore, despite carbon dioxide (CO2) emissions' significance as a global issue, the literature has paid limited attention to this area. In this report, we propose EcoLight, a reward shaping scheme for reinforcement learning algorithms that not only reduces CO2 emissions but also achieves competitive results in metrics such as travel time. We compare the performance of tabular Q-Learning, DQN, SARSA, and A2C algorithms using metrics such as travel time, CO2 emissions, waiting time, and stopped time. Our evaluation considers multiple scenarios that encompass a range of road users (trucks, buses, cars) with varying pollution levels.
\end{abstract}

\section{Introduction}
Responsive traffic lights prevent people from spending unnecessary and significant time and money on roadways. Forbes reports that traffic congestion costs the United States \$124 billion annually, while up to 1\% of the European Union’s GDP is lost due to traffic congestion. In major cities like Los Angeles, New York, and San Francisco, residents spend an average of 3-4 days each year stuck in traffic, wasting \$10 billion in fuel and time in 2017 alone \cite{INRIXglobal}. Land traffic emissions are responsible for one-third of pollution-related mortalities in North America, and air pollution results in approximately 3 million deaths globally each year \cite{world2016ambient}. Clearly, optimizing traffic flow is a critical issue, and improving traffic light control at intersections is a vital sub-problem. Suboptimal traffic control signals result in over 100 hours of additional driving time each year for big-city drivers in the United States. However, studies like \cite{tang2018research} demonstrate that improved traffic control signals can significantly reduce delays, with Hangzhou, China's signal optimization resulting in an average time savings of 4.6 minutes per vehicle and a 15.3\% reduction in delays.


A number of approaches have been proposed from simple fixed-time cycles to adaptive traffic control systems. Recently, the state-of-the-art results in traffic optimization were achieved by using Reinforcement Learning (RL) algorithms that can adjust according to traffic conditions. In much of past work, reward has been defined as an ad hoc weighted linear combination of numerous traffic measures \cite{agand2022online}.  In order to take into account more aspects of the traffic conditions, recent RL techniques incorporate sophisticated states such as images from cameras. 
This added complexity may result in a less efficient learning process without a considerable improvement in performance.

The authors in \cite{bichiou2018developing} use automated vehicles (AVs)  as an option to reduce the delay, which has CO2 reduction as a byproduct. Another study also shows how AVs can lead to significant progress towards emission reduction \cite{pribyl2020addressing}.  The study carried out by \cite{jereb2021methodology} addresses challenges of the generation of CO2 caused by urban transportation. They determined the amount of CO2 generated according to the type of vehicle, traffic flow, traffic light signal schedule, and vehicle velocity. Additionally, they look at the impact of establishing effective traffic flow management in various scenarios, demonstrating that the majority of CO2 was produced during waiting and accelerating phases in front of traffic lights as opposed to running phases through intersections.

To the best of our knowledge, our paper is the first attempt of intelligent traffic control that directly targets CO2 emissions reductions in a complex setting which includes different types of vehicles. 
To this end, we propose a reward shaping scheme that weighs different classes of road users such as cars, trucks and buses differently. 
This additional hyper-parameter allows us to adjust to different scenarios and real world objectives.  \prv{To avoid instability, the weights are initialized according to the emission class of the vehicles.}

\section{Related works}
A number of approaches have been proposed for constructing traffic light control policies. For instance,  a fixed-time cycle-based traffic signal controller  chooses the next phase by displaying it in an ordered sequence known as a cycle with each phase has a fixed, potentially unique duration. 
Researchers have long attempted to build new traffic signal controllers that can adjust to changing traffic conditions, despite the fixed-time controller's widespread use. 


\subsection{Non-Learning Traffic Signal Controllers}
Fixed-time control, actuated approaches, and selection-based adaptive control systems all rely largely on human understanding since they require manually generated traffic signal designs or regulations.  A few non-learning approaches are as follows:

\subsubsection{ Uniform (fixed-time)} A simple cycle-based, uniform phase length traffic signal controller to which other controllers can be compared as a baseline. The uniform controller's only hyper-parameter is the green duration $g$, which sets the same duration for all green phases; the next phase is determined by a cycle.

\subsubsection{ Websters}  Using phase flow data, Webster's method creates a cycle-based, fixed phase length traffic light controllers \cite{ho2016generative}. The authors propose an adaptive controller that collects data for $W$ seconds before using Webster's technique to determine the cycle and green phase duration for the following $W$ seconds. Webster recommends minimising travel time by concentrating solely on the busiest intersections and assuming a constant traffic arrival rate.

\subsubsection{Max-pressure} 
This algorithm treats traffic lanes as if they were  material in a pipe, applying control to maximise pressure relief between list of vehicles in incoming ($L_{in}$) and out-going ($L_{out}$) lanes  \cite{varaiya2013max}. Max-pressure greedily chooses the green phase ($p=1$) with maximum pressure as $\max_{p} \sum_{j\in L_{in}, p_i=1} N_j - \sum_{j\in L_{out}, p_i = 1}  N_j$ where $N_j$ is the number of vehicles in lane $j$, $p = [p_1, p_2,\cdots, p_{L_{in}}]$, and $p_i$ is a binary value that indicates the phase for the $i$-th incoming and out-going lanes.  

\subsubsection{Self Organizing Traffic Lights (SOTL)} Instead of optimizing traffic lights for a particular density and configuration of traffic, SOTL propose an adaptive feasible alternative to reflect changes in the traffic conditions \cite{goel2017self}. 


\subsection{Learning-based Traffic Signal Controllers} 
To decide on traffic signal strategies, learning-based approaches rely on observed data rather than human knowledge.
Tabular Q-Learning (QT) is limited in large state spaces due to the storage requirements of the value table \cite{el2010towards}. Authors in \cite{zhao2017novel} use Deep Q-network (DQN) that approximate the Q-function with a neural network. It has two hidden layers with exponential linear unit (ELU) and a linear output layer. The input is the local intersection state at time $t$. Authors in \cite{kekuda2021reinforcement} use State-Action-Reward-State-Action (SARSA)  as a low-cost real-time RL algorithm  to minimize congestion in networks. Authors in \cite{chu2019multi} use multi-agent advantage Actor-Critic (A2C) deep RL algorithm that improves observability and reduces the learning difficulty of each local agent. Several other deep RL algorithms consider the interactions between different intersections, such as MPlight \cite{chen2020toward}, MADQN \cite{rasheed2022deep} and others \cite{shabestary2022adaptive}.

\section{Method}

\begin{figure}[t!]
\begin{center}
\centerline{\includegraphics[width=0.75\linewidth,trim={2cm  2cm 3cm 3cm },clip]{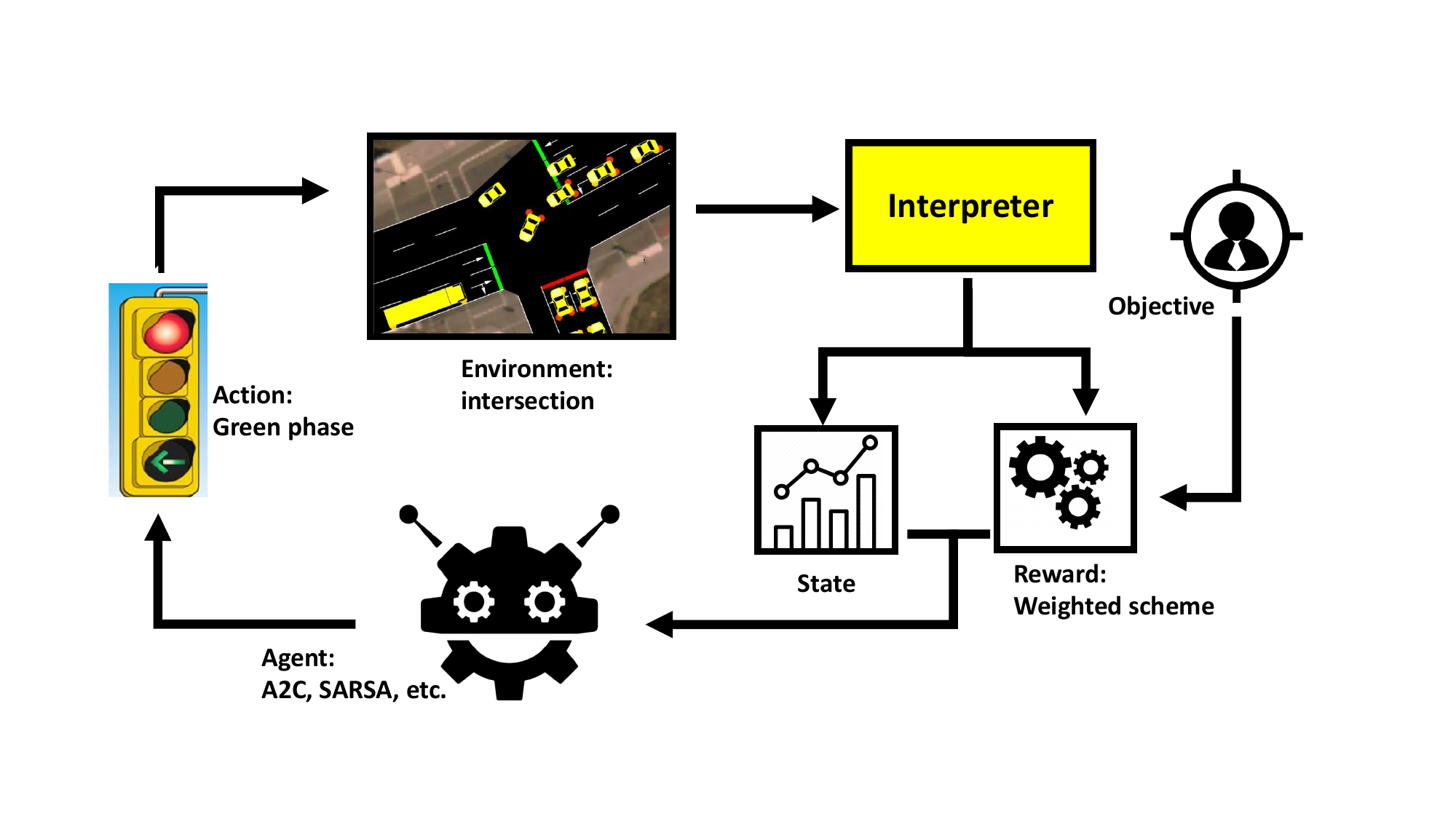}}
\caption{Ecolight RL structure}
 \label{fig:str}
\end{center}
\vskip -0.2in
\end{figure}

This section will describe the design process and reward shaping scheme  for prioritised traffic light control. We also provide guidelines for choosing the weights used in the reward function.  The structure is shown in Fig. \ref{fig:str}, where the interpreter box abstracts away any perception system that can provide the state of the environment and the reward which is based on the state.   The code is available at \url{https://github.com/pagand/Eco-Light}.

\subsection{Agent design}
\subsubsection{\textbf{State}}
The proposed state observation ($S_t$) includes the most recent phase ($p$), the lane density ($D$), queue length ($Q$), and the average vehicle type for incoming lanes ($C$) at a intersection at time $t$. The state space $S_t \in ( \mathbb{R}^{3L_{in}}\times  \mathbb{B}^{|p|+1})$ is defined as $S_t = [p,D,Q,C]$, where $B,\mathbb{R}$ are binary and real numbers.  The intersection has three valid flow and corresponding phases, north-to-south (N-S, $p=[1,0,0]$), east-to-west (E-W, $p=[0,1,0]$), and north-to-east/south-to-west (N-E, $p=[0,0,1]$). There is another phase ($p=[0,0,0]$) that encodes the all-red clearance phase. Note that the flow from north to west and south to east are allowed in phase (N-S), and  the reverse is allowed in (W-E). The other quantities are computed for each incoming lane as  follows: 
\begin{equation}
D_j = \frac{N_j}{\bar{L}G}, ~~ Q_j = \frac{N_{Hj}}{D_j}, ~~ C_j = \frac{\sum_{i\leq N_j} \mathcal{E}_{ij} }{N_j\mathcal{E}_{max}},
\end{equation}

\noindent where $G$ is the average length of vehicles plus the minimum gap between stationary vehicles. Also, $\bar{L}$ is the average length of the lanes. $\mathcal{E}_{ij}$ and $\mathcal{E}_{max}$ are the $i$-th vehicle emission class in lane $j$ and also the lanes' maximum emission which is the most inefficient vehicle class that is permitted, respectively. The queue length is defined to be the concatenation of the normalized number of vehicles traveling less than 5 km/h, also known as the halting state ($N_{Hj}$) for each incoming lane. Furthermore, the average vehicle type is defined based on the vehicles' normal emission class in each incoming lane.

\subsubsection{\textbf{Action}}
The proposed action space for the traffic signal controller determines the next green phase. Given the policy ($\pi$), the agent selects one action from a discrete set which is one of the many possible green phases. After a green phase has been selected, it is enacted for a duration equal to the minimum green phase $g_{min}$ and it can remain unchanged up to $g_{max}$.

\subsubsection{\textbf{Reward}}
We consider three different rewards: queue length ($r_q = -(\sum_{j\in L_{in}} N_{Hj} )^2$), waiting time ($r_w=0.01 \sum_{j\in L_{in}}( T_{j,t} -  T_{j,t-1})$), and pressure ($r_p = -|\sum_{j\in L_{in}} N_j-\sum_{j\in L_{out}} N_j|$), 
where $T_{j,t}$ is the overall waiting time of lane $j$ in step $t$. The vehicle is assumed to be waiting if they are in halting mode. \pa{ The behavior of an agent in its surroundings can be depicted as a Markov Decision Process (MDP).  The objective of the agent is to choose actions that maximize the "return" $R_t = \sum_{t=0}^T\gamma^t r$ where $r \in \{r_q, r_w, r_p\}$, $T$ is the time step at which the simulation terminates and $\gamma \in [0, 1)$ is the discount factor that determines the trade-off between the importance of immediate and delayed rewards.
}

 \begin{figure}[t]
\begin{center}
\centerline{\includegraphics[width=0.7\linewidth,trim={4cm  4cm 12cm 10cm },clip]{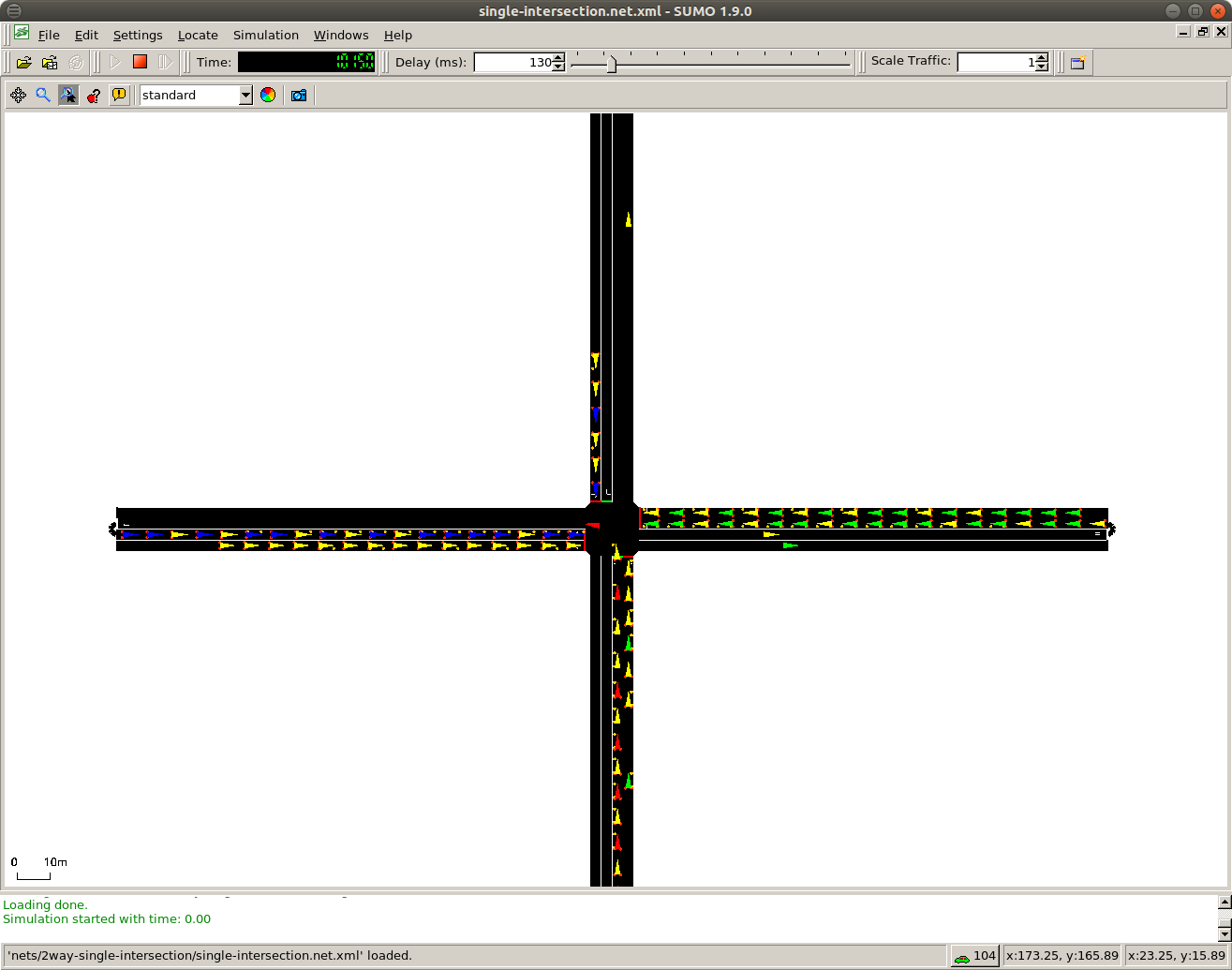}}
\caption{Two way single intersection with through, left, and right option in each lane with different road user: yellow for car, blue for truck, green for bus, red for light truck
 }
 \label{fig:ss}
\end{center}
\vskip -0.2in
\end{figure}

\subsubsection{\textbf{RL agent}}

\pa{For the agent, we consider QT, DQN and SARSA as value-based methods and A2C as a policy-based method. A drawback of neural networks that learn the Q-function directly is that they cannot independently estimate the value of a state and action \cite{hammond2020forest,agand2023fuel}. QT is an off-policy technique that utilizes the greedy approach to learn the Q-value, whereas SARSA is on-policy and learns the Q-value based on the action taken by the current policy. The loss function is:
\begin{equation}
L_{QT} = \sum_{(s,a,r,s')\sim U(B)} ([r+\gamma \max_a Q(s',a;\theta_i)]- Q(s,a;\theta_t))^2
    \label{eq:ql}
\end{equation}
where $B$ is the experience-replay memory buffer from which experiences are uniform randomly sampled and $\theta_t, \theta_i$ are the target network parameters and frozen parameters for evaluation, respectively. In SARSA algorithm,   as target value for  state-action pair is dependent on the next action the loss function is defined as:
\begin{equation}
L_{S} = \sum_{(s,a,r,s',a')\sim U(B)} ([r+\gamma Q(s',a';\theta_i)]- Q(s,a,\theta_t))^2
\end{equation}
A2C algorithm is a temporal difference (TD) variation of the policy gradient method \cite{mnih2016asynchronous,agand2022human}. It consists of two networks: the Actor network, which determines the appropriate action to take, and the Critic network, which evaluates the action's effectiveness and advises the Actor on how to improve. The losses for the Actor and Critic networks can be computed as follows:
\begin{equation}\begin{split}
L_{A,actor} =& -\log \pi(a_t|s_t;\theta_a) \tilde{A}(s_t;\theta_c)\\
L_{A,critic} =& \tilde{A}(s_t;\theta_c)^2\\
\end{split}
\end{equation}
 where $\tilde{A}(s_t;\theta_c) = r+\gamma V(s_{t+1}|s_t,a_t;\theta_c)-V(s_t;\theta_c)$ is the approximate advantage function,  $V(s_t) = \mathbb{E}[R_t|s_t]$ is the value function of state $s_t$, and $\theta_c, \theta_a$ are the critic and actor network weights, respectively.}

\subsection{Reward shaping}
In the reward shaping, we propose the weighted version of the reward functions. These weights prioritise the  vehicles with inefficient emission class, which results in a reduction of total generated carbon dioxide. The following relation, will present the weighted version rewards 
for queue length ($r_{wq}$), waiting time ($r_{ww}$), and pressure ($r_{wp}$).

\begin{equation}
\begin{split}
    r_{wq} =& -(\sum_{j\in L_{in}} N_{wHj} )^2 ,~~
    N_{wHj} = \sum_{k=1}^{N_{Hj}} W_k\\
    r_{ww} =& 0.01 \sum_{j\in L_{in}}\big( T_{wj,t} - T_{wj,t-1}\Big),~~
    T_{wj,t} = \sum_{k=1}^{N_{j}} W_k \delta_{jk,t}\\
    r_{wp} =& -\left|\sum_{j\in L_{in}} N_{wj}-\sum_{j\in L_{out}} N_{wj}\right| , ~~
    N_{wj} = \sum_{k=1}^{N_j} W_k,
    \label{eq:wreward}
\end{split}
\end{equation}

\noindent where $\delta_{jk,t}$ is the waiting time of the $k$-th vehicle in the lane $j$ at time step $t$. Also $W_k$ is a weight specific for vehicle $k$. 

\subsection{Weight selection}
We suggest three ways to determine the weights. The first way is to choose a constant value for each type of vehicle. This constant value can be optimized in different settings. The second approach is to choose the weights based on the normalized emissions of each lane, which means that all vehicles in one lane will get an equal and unique weight. This normalized number is calculated as follows:
\begin{equation}
W_j = \frac{ \mathcal{E}_j - \bar{\mathcal{E}} }{\mathcal{E}_{max}N_j},
\end{equation}
\noindent where $\mathcal{E}_j$ and  $\bar{\mathcal{E}}$ are the total and medium CO2 emissions in lane $j$, respectively.  The third way is to consider adaptive weights equal to the normalized version of the corresponding vehicle's concurrent emission \cite{agand2019adaptive}.

\subsection{Sensitivity analysis}
It is critical to know the sensitivity of any given reward's performance relative to changes in hyperparameters. \prv{Sensitivity of the weights are defined as follows:}
\begin{equation}
\gamma_{R_{w}}^{w_i} := \frac{\partial \ln ({R_{w}}^{ w_i}) }{ \partial w_i} =\frac{\partial R_{w}}{\partial w_i}\frac{w_i}{R_{w}}
\end{equation}
\prv{Hence, the sensitivity for different rewards in Eq. (\ref{eq:wreward}) are:}
\begin{equation}\begin{split}
\gamma_{R_{wq}}^{w_i} =&\frac{2w_i^2}{\sum_{j\in L_{in}} (N_{wHj})},\\
\gamma_{R_{ww}}^{w_i} =& \frac{  w_i(T_{Ji} -T_{Ji} \mathcal{I}(i,t-1)) }{\sum_{j\in L_{in}} (T_{wj,t}-T_{wj,t-1})},\\
\gamma_{R_{wp}}^{w_i}  =& w_i/\left|\sum_{j\in L_{in}} N_{wj}-\sum_{j\in L_{out}} N_{wj}\right|
    \label{eq:sen}
\end{split}
\end{equation}

\noindent where $J$ is the lane that the corresponding vehicle $i$ is in, and $\mathcal{I}(i,t)$ is an indicator, which is zero only if the vehicle $i$ was moving at time $t$.

\section{Experiments}

\subsection{Setup}
Software that is used includes SUMO v1.9.2 \cite{krajzewicz2012recent}, Pytorch v1.8.1, Stable-Baselines, Stable-Baseline3 (SB3) v1.0, and Python v3.7. We use the Adam optimizer for gradient-based optimization wherever applicable. The scenario in the SUMO environment is shown in Fig. \ref{fig:ss}. Travel time is computed as:
\begin{equation}
\mathcal{T}=\frac{\bar{L} \sum_{j \in L_{in} \cup L_{out}} N_j}{\sum_{j \in L_{in} \cup L_{out}} (\bar{V}_j N_j)},
    \label{eq:traveltime}
\end{equation}
\noindent where $\bar{V}_j$ is the average velocity of vehicles in lane $j$. As mentioned before, we consider an intersection with 4 directions, each with 2 incoming and 2 outgoing lanes, and a length of 150 m. The simulation sample time interval is set to 5 seconds. The total duration of all simulations is 100,000 time steps (approximately 6 days). For green phase, we consider $g_{min}=10, g_{max}=50$ time steps. There are three major traffic flows that are injected into the original traffic at time steps 25,000, 50,000, and 75,000 of the simulation.

For the fixed-time method, we consider the fixed green time of north-south and east-west to be equal to 42 steps. For learning hyper-parameters, we use grid-based search. For QT, we consider learning rate of $\alpha = 0.1$, discount factor of $\gamma = 0.99$, exploration rate of $\epsilon =0.05$, final exploration of  $\epsilon_{min}=0.005$, weight decay of $1$, and an epsilon-greedy policy. For the DQN method, we used an MLP for the policy network of Stable-Baseline3 with $\alpha = 0.01$, target update interval $100$, $\epsilon =0.05$, and $\epsilon_{min}=0.01$. For A2C, we use the synchronous deterministic variant of A3C \cite{mnih2016asynchronous} with MLP policy network and Stable-Baselines for implementation. To avoid using a replay buffer, it employs many workers with $\alpha=0.001$ and uses Kronecker-Factored Trust Region (ACKTR) method \cite{wu2017scalable}. ACKTR is a second-order optimization method that increase sample efficiency and scalability by utilizing trust region for more consistent improvement, and distributed Kronecker factorization for approximation. Finally, for SARSA, we use an online algorithm with Fourier order 7 for linear function approximation and TD-$\lambda$ to compute the return \cite{van2016true} with $\lambda = 0.95$, $\alpha = 0.001$, $\gamma = 0.95$, and $\epsilon = 0.01$.

\subsection{Hyper-parameter tuning and  robustness of weights}
\begin{figure}[ht]
\begin{center}
\centerline{\includegraphics[width=0.95\linewidth,trim={2.5cm  0cm 2.5cm 0.5cm },clip]{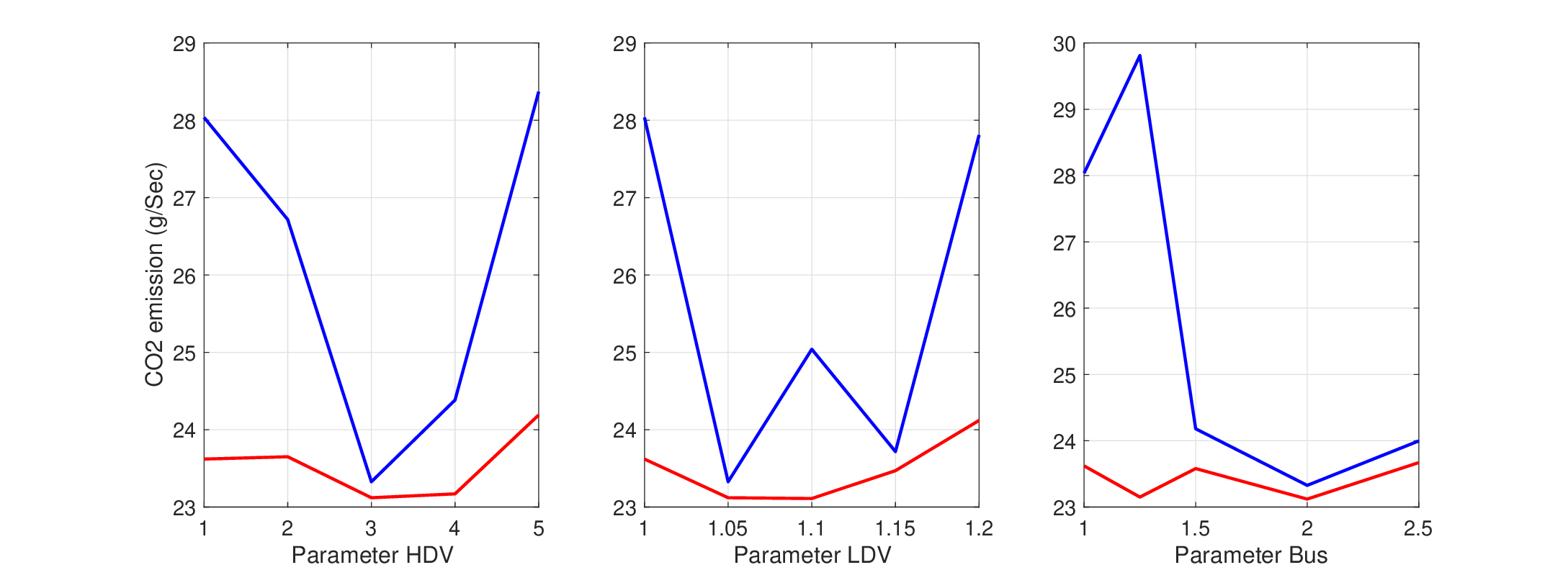}}
\caption{Tuning reward weights  $[W_k^{HDV}, W_k^{Bus}, W_k^{LDV}]$  to optimize CO2 emissions  of two different scenarios using waiting time reward, and SARSA algorithm}
 \label{fig:hyper}
\end{center}
\vskip -0.2in
\end{figure}

\prv{For hyperparameter tuning, we utilize a grid search approach to compare the CO\textsubscript{2} emissions of different reward weights. We implement two scenarios on distinct, single, 2-way intersections and consider the waiting time reward with the SARSA algorithm with a constant weight as our method. As illustrated in Fig. \ref{fig:hyper}, among the considered weights ($W_k \in \{1,2,3,4,5\}$), the optimal values are $W_k^\ast = [W_k^{HDV\ast}, W_k^{Bus\ast}, W_k^{LDV\ast}] = [3,2,1.05]$ for heavy-duty vehicle (HDV), Bus, and light-duty vehicle (LDV), respectively. This implies that an HDV priority is considered equivalent to three passenger cars in a conflicting flow.}

\begin{figure}[ht]
\begin{center}
\centerline{\includegraphics[width=0.95\linewidth,trim={0.8cm  0.2cm 0.6cm 1.5cm },clip]{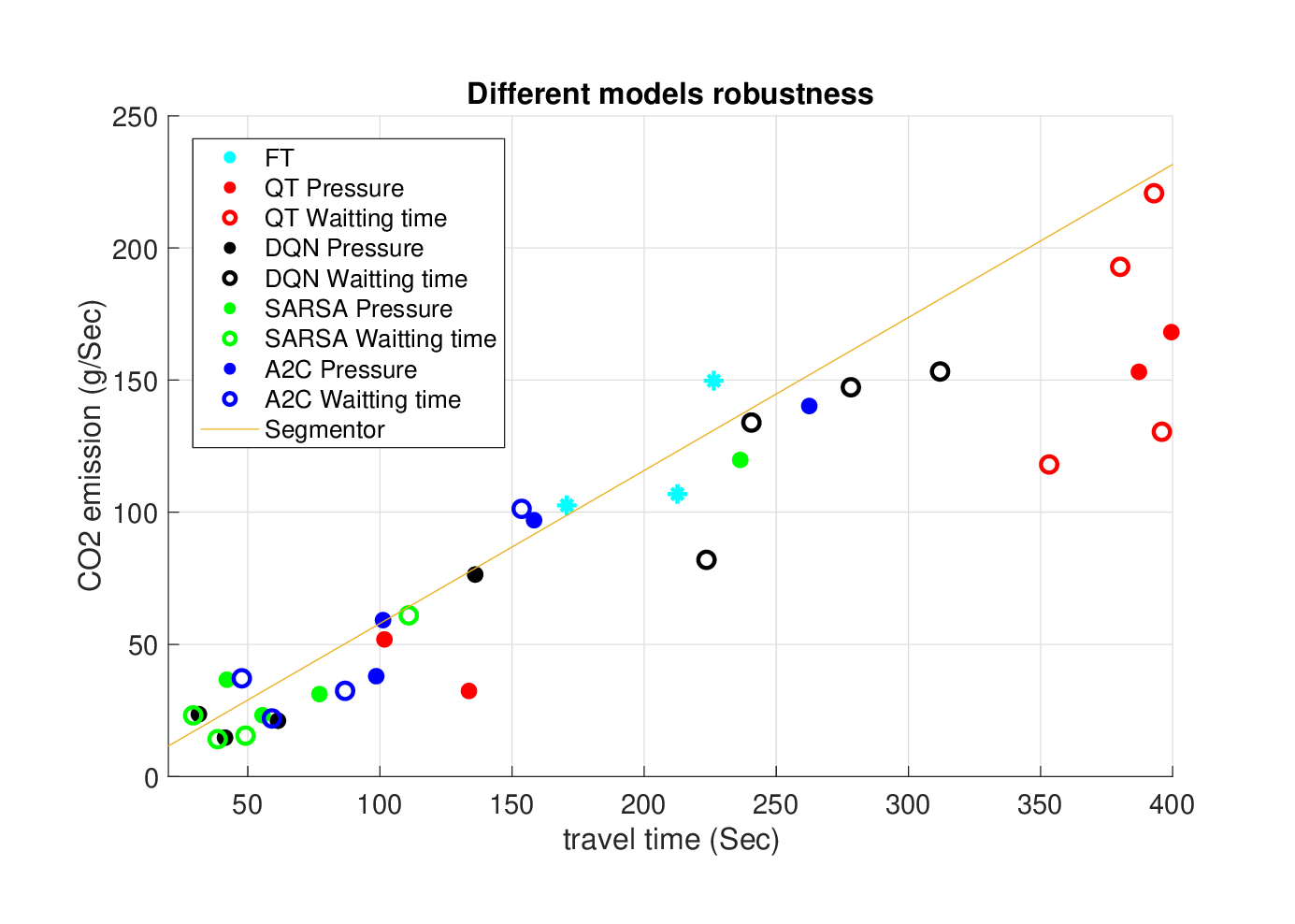}}
\caption{Comparison between different weighted rewards of waiting time, pressure,  and fixed-time in different settings}
 \label{fig:comp}
\end{center}
\vskip -0.2in
\end{figure}

\prv{To assess the robustness of weight values, we explore four different cases ($W_k \in \{W_k^\ast, 1.1W_k^\ast-0.1, 0.9W_k^\ast+0.1, 1.7W_k^\ast-0.7\}$). Fig. \ref{fig:comp} compares their effects on the weighted waiting time and pressure rewards applied under different RL agents. In general, the pressure reward approach exhibits superior performance, and adjusting the weights shifts the points above or below the separator. This observation emphasizes the importance of carefully selecting weights to strike a balance between travel time and CO2 emissions. The results from the sensitivity analysis in Eq. (\ref{eq:sen}) are consistent, as waiting time consistently holds a larger value in the numerator compared to the pressure reward, resulting in an overall less robust approach. Additionally, it is evident that the queue length has a quadratic relation to the initial weight, rendering it the least robust approach to weight perturbation.}

\begin{table*}[ht!]
\caption{Comparing extended experiments inspired from Pressure DQN  \cite{wei2019presslight} , queue length A2C \cite{chaudhuri2022comparative}, waiting time DQN  \cite{alegre2021quantifying}, queue length SARSA  \cite{kekuda2021reinforcement}, and waiting time SARSA \cite{sumorl}}
\begin{center}
\begin{tabular}{ccccccccccccc} 
\multirow{2}{*}{metric}&\multirow{2}{*}{ratio}&\multirow{2}{*}{type}&\multirow{2}{*}{fixed-time}&\multicolumn{3}{c}{ ------- Waiting  time -------}&\multicolumn{3}{c}{ ------- Queue length ------- }&\multicolumn{3}{c}{------- Pressure -------}\\
   & &&&DQN&A2C&SARSA&DQN&A2C&SARSA&DQN&A2C&SARSA\\\toprule
 \multirow{6}{*}{Travel time}&   \multirow{2}{*}{0\%}&Baseline&  \multirow{2}{*}{170.73}&249.54&49.32&\bf{29.75}&78.09&90.06&\bf{32.51}&\bf{32.12}&126.13&40.18\\
    &&Ours&&240.59&47.75&\bf{29.34}&126.95&119.54&35.37&\bf{31.50}&158.34&42.07\\
    \cline{2-13}
    &\multirow{2}{*}{10\%}&Baseline&  \multirow{2}{*}{212.58}&385.87&56.10&\bf{38.57}&188.63&129.36&43.02&46.52&142.15&52.70\\
    &&Ours&&223.58&59.14&\bf{38.61}&317.28&106.18&43.06&\bf{41.39}&98.61&55.57\\
    \cline{2-13}
    &\multirow{2}{*}{40\%}&Baseline&  \multirow{2}{*}{226.34}&248.39&162.40&125.67&322.46&224.11&157.38&137.10&248.43&210.06\\
    &&Ours&&311.99&153.64&\bf{110.91}&295.18&229.43&164.34&136.07&262.48&236.36\\
    \bottomrule 
    \multirow{6}{*}{CO2 emission}&\multirow{2}{*}{0\%}&Baseline&  \multirow{2}{*}{102.60}&134.79&38.40&\bf{23.62}&55.88&64.53&26.93&\bf{24.35}&77.68&33.92\\
    &&Ours&&133.93&37.17&\bf{23.14}&88.67&77.49&29.77&\bf{23.58}&96.98&36.65\\
    \cline{2-13}
    &\multirow{2}{*}{10\%}&Baseline&  \multirow{2}{*}{106.91}&165.08&21.42&\bf{14.21}&79.19&45.56&18.08&17.48&53.37&22.26\\
    &&Ours&&81.99&21.98&\bf{14.16}&157.90&38.63&16.53&\bf{14.72}&37.97&23.21\\
    \cline{2-13}
    &\multirow{2}{*}{40\%}&Baseline&  \multirow{2}{*}{149.76}&152.19&113.48&84.11&167.57&145.45&111.26&93.31&135.85&128.35\\
    &&Ours&&153.25&101.29&\bf{69.98}&143.61&123.43& 84.96&76.43&140.19&119.79\\
    \bottomrule 
    \multirow{6}{*}{Waiting time}&\multirow{2}{*}{0\%}&Baseline&  \multirow{2}{*}{9351}&22983&454.16&\bf{82.51}&8930&1444&188.08&457.78&1998&326.80\\
    &&Ours&&43044&407.82&\bf{79.28}&44981&1995&199.07&274.51&3576&433.08\\
    \cline{2-13}
    &\multirow{2}{*}{10\%}&Baseline&  \multirow{2}{*}{12114}&134473&178.87& \bf{41.27}&20294&1564&118.59&329.04&2054&246.67\\
    &&Ours&&7489&201.49&\bf{41.55}&93558&738.65&98.15&137.74&778.38&286.28\\
    \cline{2-13}
    &\multirow{2}{*}{40\%}&Baseline&  \multirow{2}{*}{15337}&11525&2371&1091&21178&5365&7442&2417&5025&15665\\
    &&Ours&&22041&2117&\bf{788.06}&26812&4878&5138&1840&6544&11109\\
    \bottomrule 
     \multirow{6}{*}{Stopped time}&\multirow{2}{*}{0\%}&Baseline&  \multirow{2}{*}{37.40}&49.15&11.58&\bf{6.24}&18.94&21.24&7.43&\bf{6.89}&26.13&9.98\\
    &&Ours&&48.66&11.12&\bf{6.07}&31.59&26.06&8.45&\bf{6.67}&33.59&10.93\\
    \cline{2-13}
    &\multirow{2}{*}{10\%}&Baseline&  \multirow{2}{*}{26.22}&59.18&5.47&\bf{3.18}&23.20&14.02&\bf{3.99}&\bf{3.78}&15.58&4.96\\
    &&Ours&&28.13&5.75&\bf{3.20}&49.54&11.61&\bf{3.89}&\bf{3.23}&10.79&5.92\\
    \cline{2-13}
    &\multirow{2}{*}{40\%}&Baseline&  \multirow{2}{*}{32.70}&40.37&24.24&17.55&45.43&31.62&22.08&20.25&31.95&30.16\\
    &&Ours&&43.85&23.14&\bf{15.57}&40.37&30.27&23.80&18.10&33.76&35.77\\
    \bottomrule 
\end{tabular}
\end{center}
\label{tab:all}
\end{table*}

\subsection{Comparison}
 \begin{figure}[t]
\begin{center}
\centerline{\includegraphics[width=1\linewidth,trim={0.1cm  0cm 0cm 0cm },clip]{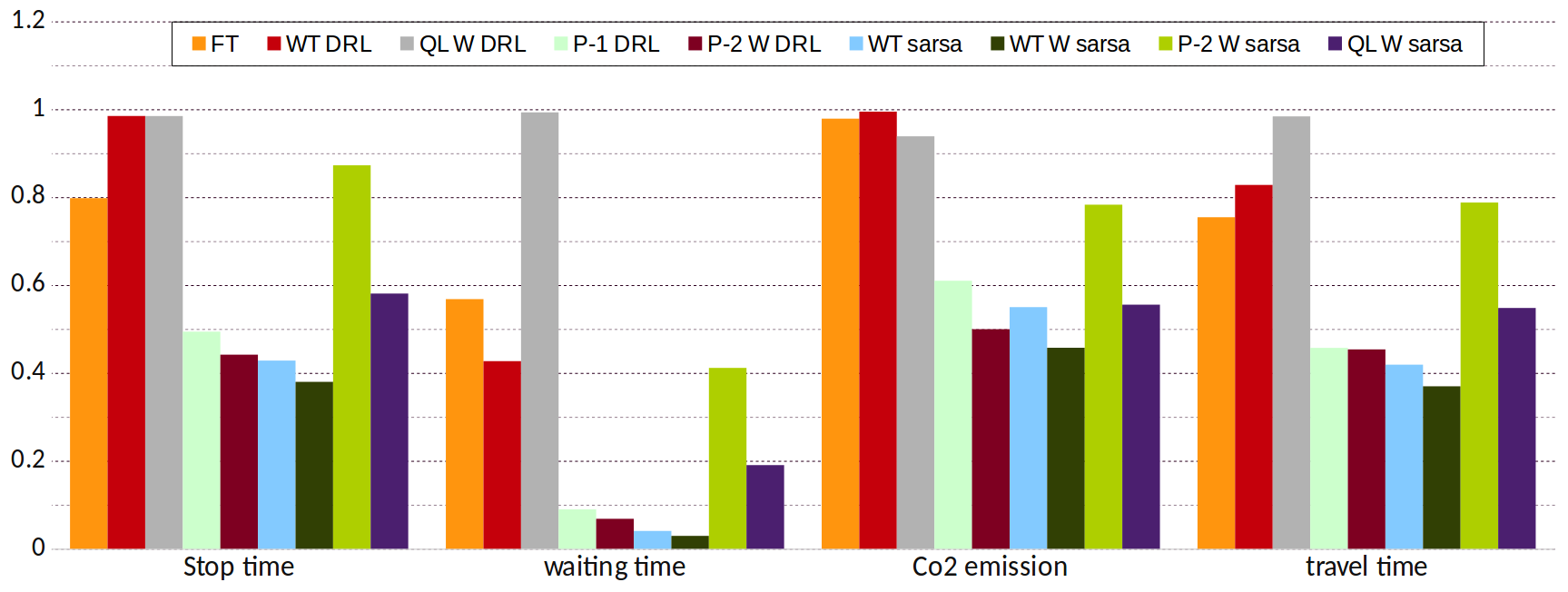}}
\caption{Comparison for top performance approaches with respect to different traffic criteria
 }
 \label{fig:bar}
\end{center}
\vskip -0.2in
\end{figure}

According to Fig. \ref{fig:bar}, which compares the top-performing approaches, QT is unable to perform well in sophisticated traffic situations because the number of states is not sufficient to fully represent the traffic complexity. SARSA and A2C can outperform DQN in most cases. Using a weighted reward function with a suitable RL approach, such as SARSA, could decrease CO2 emissions by up to 50\%. FT (fixed-time), WT (waiting time), QL (queue length), P-1/2 (Pressure reward using first two approaches of weight selection), and W (weighted versions of the rewards) are compared for deep Q-Learning (DRL) and SARSA approaches.

For quantitative comparison, the average metrics of the last 10,000 steps using different rewards and RL algorithm  are presented for unweighted rewards (baseline) and weighted rewards (ours) in Table \ref{tab:all}. We spawned a flow of different road users, with a ratio of 0 (only normal vehicles), 10\%, and 40\% (for every 100 vehicles, 40 are normal and the rest are HDV/Bus/LDV) in simulations to see effectiveness of each approach in different scenarios. Since some vehicles are less efficient in terms of fuel consumption, they should be granted a higher priority compared to normal vehicles. When comparing travel time vs. CO2 emissions for a higher ratio, only DQN with waiting time reward had the best performance among all. However, for lower ratios, DQN with pressure reward could also be among the top results. Additionally, the difference between weighted and unweighted rewards becomes noticeable for larger ratios. For the waiting time comparison, regardless of the ratio, the waiting time approach with the SARSA algorithm provides the best performance.

\section{Discussion}
\begin{figure}[t]
\begin{center}
\centerline{\includegraphics[width=1\linewidth,trim={2.5cm  2.5cm 2.5cm 2.1cm },clip]{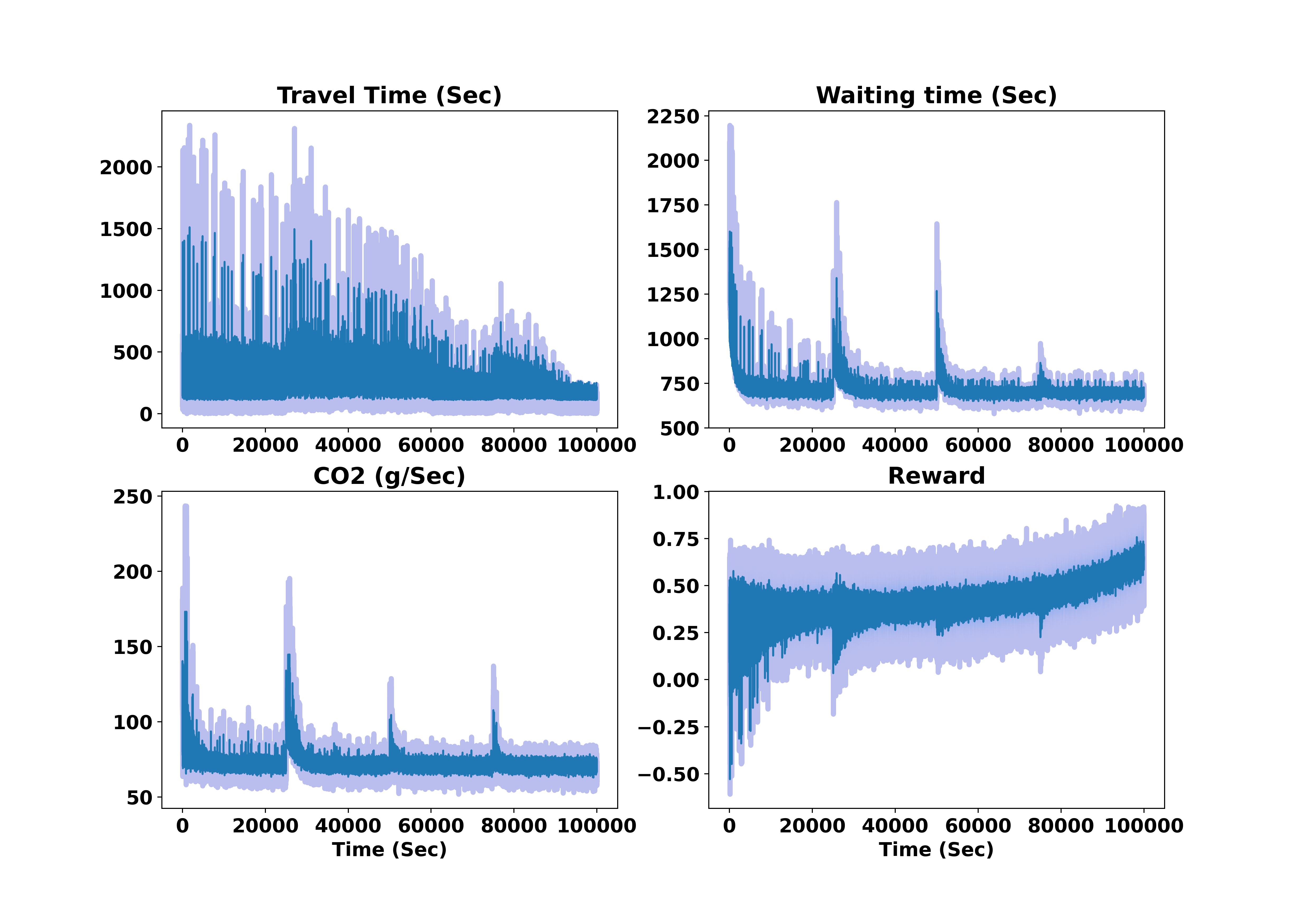}}
\caption{Travel time, CO2 emission, waiting time and reward for weighted waiting time, SARSA algorithm for ten runs. }
 \label{fig:all}
\end{center}
\vskip -0.2in
\end{figure}

For further elaboration, the profiles of the weighted waiting time with SARSA for different metrics are shown in Fig. \ref{fig:all}. \prv{There are three peak traffic injection at time 2500, 5000, and 7500 which explains the spikes in the results.}  Note that these plots include the training phase, as we wanted to show how quickly the agent is able to converge. As we can see, after almost 80,000 steps, despite changes in the traffic flow, the profiles of all traffic elements have negligible fluctuations. This shows that the policy network with weighted reward functions was able to function responsively and efficiently. It should be noted that the performance of these results in implementation is limited by the accuracy of the vehicle class detector. Therefore, integrating the results with a computer vision system is suggested to implement this approach in real-world applications. \prv{
While in our approach we advocate prioritizing inefficient vehicles, it is essential to clarify that this should not be perceived as an incentive. Instead, certain penalties must be enforced by the city to discourage inefficient vehicle usage. } The flow of neighboring intersections is neglected in the decision-making of EcoLight. As a future direction, implementing a network approach for this local approach can benefit from unique properties such as green waves or avoiding spikes during engine starting for HDVs.

\section{Conclusion}
We have proposed a reward shaping scheme using weighted versions of pressure, waiting time, and queue length to address the problem of minimizing CO2 emissions at signalized intersections. To prioritize vehicles with inefficient fuel consumption or those given priority by city officials, we have considered different vehicle types at the intersection. Our approach involves introducing new weighted reward functions that reduce travel time, waiting time, and stopped time while minimizing emissions. Through robustness and sensitivity analysis, we have determined sub-optimal hyperparameters and acknowledged the trade-off between CO2 emissions and overall travel time.


\bibliographystyle{IEEEtran}
\bibliography{myRef}

\begin{thebibliography}{4}
\providecommand{\natexlab}[1]{#1}
\providecommand{\url}[1]{\texttt{#1}}
\providecommand{\urlprefix}{URL }
\expandafter\ifx\csname urlstyle\endcsname\relax
  \providecommand{\doi}[1]{doi:\discretionary{}{}{}#1}\else
  \providecommand{\doi}{doi:\discretionary{}{}{}\begingroup
  \urlstyle{rm}\Url}\fi

\bibitem[{Able(1956)}]{Abl:56}
Able, B. (1956).
\newblock Nucleic acid content of microscope.
\newblock \emph{Nature}, 135, 7--9.

\bibitem[{Able et~al.(1954)Able, Tagg, and Rush}]{AbTaRu:54}
Able, B., Tagg, R., and Rush, M. (1954).
\newblock Enzyme-catalyzed cellular transanimations.
\newblock In A.~Round (ed.), \emph{Advances in Enzymology}, volume~2, 125--247.
  Academic Press, New York, 3rd edition.

\bibitem[{Keohane(1958)}]{Keo:58}
Keohane, R. (1958).
\newblock \emph{Power and Interdependence: World Politics in Transitions}.
\newblock Little, Brown \& Co., Boston.

\bibitem[{Powers(1985)}]{Pow:85}
Powers, T. (1985).
\newblock Is there a way out?
\newblock \emph{Harpers}, 35--47.

\end{thebibliography}


\begin{thebibliography}{10}
\providecommand{\url}[1]{#1}
\csname url@rmstyle\endcsname
\providecommand{\newblock}{\relax}
\providecommand{\bibinfo}[2]{#2}
\providecommand\BIBentrySTDinterwordspacing{\spaceskip=0pt\relax}
\providecommand\BIBentryALTinterwordstretchfactor{4}
\providecommand\BIBentryALTinterwordspacing{\spaceskip=\fontdimen2\font plus
\BIBentryALTinterwordstretchfactor\fontdimen3\font minus
  \fontdimen4\font\relax}
\providecommand\BIBforeignlanguage[2]{{%
\expandafter\ifx\csname l@#1\endcsname\relax
\typeout{** WARNING: IEEEtran.bst: No hyphenation pattern has been}%
\typeout{** loaded for the language `#1'. Using the pattern for}%
\typeout{** the default language instead.}%
\else
\language=\csname l@#1\endcsname
\fi
#2}}

\bibitem{INRIXglobal}
G.~Cookson, ``Inrix global traffic scorecard,'' \emph{Tech.Rep.}, 2018.

\bibitem{world2016ambient}
W.~H. Organization \emph{et~al.}, ``Ambient air pollution: A global assessment
  of exposure and burden of disease,'' 2016.

\bibitem{tang2018research}
Y.~Tang, Y.~Xiong, W.~Yu, C.~Tian, and Z.~Bao, ``Research on the development of
  intelligent transportation based on smart city,'' in \emph{2018 3rd
  international conference on control, automation and artificial intelligence
  (CAAI 2018)}.\hskip 1em plus 0.5em minus 0.4em\relax Atlantis Press, 2018,
  pp. 92--95.

\bibitem{agand2022online}
P.~Agand, M.~Chen, and H.~D. Taghirad, ``Online probabilistic model
  identification using adaptive recursive mcmc,'' in \emph{2023 International
  Joint Conference on Neural Networks (IJCNN)}.\hskip 1em plus 0.5em minus
  0.4em\relax IEEE, 2023, pp. 1--6.

\bibitem{bichiou2018developing}
Y.~Bichiou and H.~A. Rakha, ``Developing an optimal intersection control system
  for automated connected vehicles,'' \emph{IEEE Transactions on Intelligent
  Transportation Systems}, vol.~20, no.~5, pp. 1908--1916, 2018.

\bibitem{pribyl2020addressing}
O.~Pribyl, R.~Blokpoel, and M.~Matowicki, ``Addressing eu climate targets:
  Reducing co2 emissions using cooperative and automated vehicles,''
  \emph{Transportation Research Part D: Transport and Environment}, vol.~86, p.
  102437, 2020.

\bibitem{jereb2021methodology}
B.~Jereb, O.~Stopka, and T.~Skr{\'u}can{\`y}, ``Methodology for estimating the
  effect of traffic flow management on fuel consumption and co2 production: a
  case study of celje, slovenia,'' \emph{Energies}, vol.~14, no.~6, p. 1673,
  2021.

\bibitem{ho2016generative}
J.~Ho and S.~Ermon, ``Generative adversarial imitation learning,''
  \emph{Advances in neural information processing systems}, vol.~29, pp.
  4565--4573, 2016.

\bibitem{varaiya2013max}
P.~Varaiya, ``The max-pressure controller for arbitrary networks of signalized
  intersections,'' in \emph{Advances in Dynamic Network Modeling in Complex
  Transportation Systems}.\hskip 1em plus 0.5em minus 0.4em\relax Springer,
  2013, pp. 27--66.

\bibitem{goel2017self}
S.~Goel, S.~F. Bush, and C.~Gershenson, ``Self-organization in traffic lights:
  Evolution of signal control with advances in sensors and communications,''
  \emph{arXiv preprint arXiv:1708.07188}, 2017.

\bibitem{el2010towards}
S.~El-Tantawy and B.~Abdulhai, ``Towards multi-agent reinforcement learning for
  integrated network of optimal traffic controllers (marlin-otc),''
  \emph{Transportation Letters}, vol.~2, no.~2, pp. 89--110, 2010.

\bibitem{zhao2017novel}
Y.~Zhao, H.~Gao, S.~Wang, and F.-Y. Wang, ``A novel approach for traffic signal
  control: A recommendation perspective,'' \emph{IEEE Intelligent
  Transportation Systems Magazine}, vol.~9, no.~3, pp. 127--135, 2017.

\bibitem{kekuda2021reinforcement}
A.~Kekuda, R.~Anirudh, and M.~Krishnan, ``Reinforcement learning based
  intelligent traffic signal control using n-step sarsa,'' in \emph{2021
  International Conference on Artificial Intelligence and Smart Systems
  (ICAIS)}.\hskip 1em plus 0.5em minus 0.4em\relax IEEE, 2021, pp. 379--384.

\bibitem{chu2019multi}
T.~Chu, J.~Wang, L.~Codec{\`a}, and Z.~Li, ``Multi-agent deep reinforcement
  learning for large-scale traffic signal control,'' \emph{IEEE Transactions on
  Intelligent Transportation Systems}, vol.~21, no.~3, pp. 1086--1095, 2019.

\bibitem{chen2020toward}
C.~Chen, H.~Wei, N.~Xu, G.~Zheng, M.~Yang, Y.~Xiong, K.~Xu, and Z.~Li, ``Toward
  a thousand lights: Decentralized deep reinforcement learning for large-scale
  traffic signal control,'' in \emph{Proceedings of the AAAI Conference on
  Artificial Intelligence}, vol.~34, no.~04, 2020, pp. 3414--3421.

\bibitem{rasheed2022deep}
F.~Rasheed, K.-L.~A. Yau, R.~M. Noor, and Y.-W. Chong, ``Deep reinforcement
  learning for addressing disruptions in traffic light control,''
  \emph{CMC-Computers Materials \& Continua}, vol.~71, no.~2, pp. 2225--2247,
  2022.

\bibitem{shabestary2022adaptive}
S.~M.~A. Shabestary and B.~Abdulhai, ``Adaptive traffic signal control with
  deep reinforcement learning and high dimensional sensory inputs: Case study
  and comprehensive sensitivity analyses,'' \emph{IEEE Transactions on
  Intelligent Transportation Systems}, 2022.

\bibitem{hammond2020forest}
T.~Hammond, D.~J. Schaap, M.~Sabatelli, and M.~A. Wiering, ``Forest fire
  control with learning from demonstration and reinforcement learning,'' in
  \emph{2020 International Joint Conference on Neural Networks (IJCNN)}.\hskip
  1em plus 0.5em minus 0.4em\relax IEEE, 2020, pp. 1--8.

\bibitem{agand2023fuel}
P.~Agand, A.~Kennedy, T.~Harris, C.~Bae, M.~Chen, and E.~J. Park, ``Fuel
  consumption prediction for a passenger ferry using machine learning and
  in-service data: A comparative study,'' \emph{Ocean Engineering}, vol. 284,
  p. 115271, 2023.

\bibitem{mnih2016asynchronous}
V.~Mnih, A.~P. Badia, M.~Mirza, A.~Graves, T.~Lillicrap, T.~Harley, D.~Silver,
  and K.~Kavukcuoglu, ``Asynchronous methods for deep reinforcement learning,''
  in \emph{International conference on machine learning}.\hskip 1em plus 0.5em
  minus 0.4em\relax PMLR, 2016, pp. 1928--1937.

\bibitem{agand2022human}
P.~Agand, M.~Taherahmadi, A.~Lim, and M.~Chen, ``Human navigational intent
  inference with probabilistic and optimal approaches,'' in \emph{2022
  International Conference on Robotics and Automation (ICRA)}.\hskip 1em plus
  0.5em minus 0.4em\relax IEEE, 2022, pp. 8562--8568.

\bibitem{agand2019adaptive}
P.~Agand and M.~A. Shoorehdeli, ``Adaptive model learning of neural networks
  with uub stability for robot dynamic estimation,'' in \emph{2019
  International Joint Conference on Neural Networks (IJCNN)}.\hskip 1em plus
  0.5em minus 0.4em\relax IEEE, 2019, pp. 1--6.

\bibitem{krajzewicz2012recent}
D.~Krajzewicz, J.~Erdmann, M.~Behrisch, and L.~Bieker, ``Recent development and
  applications of sumo-simulation of urban mobility,'' \emph{International
  journal on advances in systems and measurements}, vol.~5, no. 3\&4, 2012.

\bibitem{wu2017scalable}
Y.~Wu, E.~Mansimov, R.~B. Grosse, S.~Liao, and J.~Ba, ``Scalable trust-region
  method for deep reinforcement learning using kronecker-factored
  approximation,'' \emph{Advances in neural information processing systems},
  vol.~30, 2017.

\bibitem{van2016true}
H.~Van~Seijen, A.~R. Mahmood, P.~M. Pilarski, M.~C. Machado, and R.~S. Sutton,
  ``True online temporal-difference learning,'' \emph{The Journal of Machine
  Learning Research}, vol.~17, no.~1, pp. 5057--5096, 2016.

\bibitem{wei2019presslight}
H.~Wei, C.~Chen, G.~Zheng, K.~Wu, V.~Gayah, K.~Xu, and Z.~Li, ``Presslight:
  Learning max pressure control to coordinate traffic signals in arterial
  network,'' in \emph{Proceedings of the 25th ACM SIGKDD International
  Conference on Knowledge Discovery \& Data Mining}, 2019, pp. 1290--1298.

\bibitem{chaudhuri2022comparative}
H.~Chaudhuri, V.~Masti, V.~Veerendranath, and S.~Natarajan, ``A comparative
  study of algorithms for intelligent traffic signal control,'' in
  \emph{Machine Learning and Autonomous Systems: Proceedings of ICMLAS
  2021}.\hskip 1em plus 0.5em minus 0.4em\relax Springer, 2022, pp. 271--287.

\bibitem{alegre2021quantifying}
L.~N. Alegre, A.~L. Bazzan, and B.~C. da~Silva, ``Quantifying the impact of
  non-stationarity in reinforcement learning-based traffic signal control,''
  \emph{PeerJ Computer Science}, vol.~7, p. e575, 2021.

\bibitem{sumorl}
L.~N. Alegre, ``{SUMO-RL},'' \url{https://github.com/LucasAlegre/sumo-rl},
  2019.

\end{thebibliography}

\end{document}